\documentclass[letterpaper]{article}

\usepackage[T1]{fontenc}

\usepackage{geometry}
\usepackage{setspace}

\usepackage[articletitle=true]{achemso}

\usepackage{graphicx}
\usepackage{float}
\newfloat{scheme}{htbp}{los}
\floatname{scheme}{Scheme}
\floatname{chart}{Chart}
\newfloat{graph}{htbp}{loh}

\usepackage{chemformula} 
\usepackage[version = 4]{mhchem} 

\usepackage{authblk}
\author[1,2]{Dipankar~Jana*}
\affil[1]{Institute for Functional Intelligent Materials, National University of Singapore, 117544, Singapore}
\affil[2]{Laboratoire National des Champs Magn\'etiques Intenses, LNCMI-EMFL, CNRS UPR3228,Univ. Grenoble Alpes, Univ. Toulouse, Univ. Toulouse 3, INSA-T, Grenoble and Toulouse, France}
\author[3]{Aljoscha~Soll}
\author[3]{Zdenek Sofer}
\affil[3]{Department of Inorganic Chemistry, University of Chemistry and Technology Prague, 16628 Prague, Czech Republic}
\author[2,4]{Milan~Orlita}
\affil[4]{Institute of Physics, Charles University, Ke Karlovu 5, Prague, 121 16, Czech Republic}
\author[2]{Clement~Faugeras}
\author[1,5]{Maciej~Koperski}
\affil[5]{Department of Materials Science and Engineering, National University of Singapore, 117575, Singapore}
\author[2,6,7]{Marek~Potemski*}
\affil[6]{Faculty of Physics, University of Warsaw, 02-093 Warsaw, Poland}
\affil[7]{CENTERA, CEZAMAT, Warsaw University of Technology, 02-822 Warsaw, Poland}

\title{Spin-flip optical excitations in van der Waals antiferromagnet CrPS$_4$}
\date{*Email: jana.d02@nus.edu.sg, marek.potemski@lncmi.cnrs.fr}

\begin{document}

\maketitle

\begin{abstract}
We investigate the near-infrared optical response of the semiconducting van der Waals antiferromagnet CrPS$_4$ and identify previously unreported spin-entangled optical resonances. The strong and anisotropic magnetic-field dependence of these resonances reflects the underlying magnetic order and confirms the biaxial antiferromagnetic nature of CrPS$_4$. From the magnetic field evolution of the optical transition, we extract key magnetic parameters, including the spin-flop ($\approx0.9$~T) and spin-saturation ($\approx8$~T) fields. These results demonstrate a potential pathway for all-optical probing of spin states in van der Waals antiferromagnets, with relevance for spin-sensitive optoelectronic and magneto-optical devices.
\end{abstract}

\section*{Keywords}

2D antiferromagnets, magnetic anisotropy, spin-entangled exciton



\section{Introduction}

Van der Waals materials hosting magnetically ordered phases provide a suitable platform to explore magnetism, ultimately in the two-dimensional limit, and enable the development of novel devices in the areas of spintronics, magnonics, and THz technologies~\cite{park20252dvanderwaals}. Among materials of current interest are semiconducting van der Waals antiferromagnets (S-vdW-AFs)~\cite{rahman2021recent}. The magnetic properties of these materials are often studied with magneto-optical methods, primarily by tracing the characteristic magnon excitations that appear in the low-energy (GHz-THz) spectral range~\cite{balkanski1987effects, loudon1968theory, wang2025review, ohlmann1961antiferromagnetic, Afanasiev2021, matthiesen2023controlling, na2026engineering}. Spin order is, however, also known to affect the electronic properties, resulting in the coupling of ions’ spins to higher-energy optical transitions~\cite{wilson2021interlayer, grzeszczyk2023strongly, jana2025deconstruction}. Such spin-entangled optical transitions are of particular interest in S-vdW-AFs in which they frequently appear in the form of surprisingly sharp optical resonances~\cite{Banda1986, Gnatchenko2011, Kang2020, Son2022}. Their presence only in the antiferromagnetic phase, as traced with temperature-dependent studies, is the very first signature of the spin-coupling effect. More can be learned with inspection of their evolution when applying the magnetic field, which effectively modifies the character of spin order in antiferromagnets~\cite{Wang2021, Dipankar2023}. Exploring spin-entangled optical resonances provides an additional tool for studying the magnetic properties of S-vdW-AFs and opens up possibilities for optical manipulation of spin ordering and magnon responses. Selective excitation of these resonances may inspire the design of fast-operating spintronic or data transmission devices~\cite{kirilyuk2010ultrafast}.

With the comprehensive study of the near-infrared optical response of CrPS$_4$ S-vdW-AF, we report the identification of yet unobserved optical resonances in this material. Our focus is directed on the properties of two optical resonances:  the ultra-narrow photoluminescence (PL) transition spotted at the energy of $\approx1.35$~eV and another relatively narrow transition observed at $\approx1.38$~eV in PL and absorption spectra. These transitions, which appear only at low temperatures, in the antiferromagnetic phase of CrPS$_4$, are shown to be critically dependent on the character of the spin order that is altered by applying the magnetic field. Our observations confirm that CrPS$_4$ is a biaxial antiferromagnet. We establish an optical method to determine its characteristic parameters, such as the spin-flop, spin-saturation fields, and \textit{g}-factor. The origin of the optical response of CrPS$_4$ in the near-infrared range is discussed in the frame of \textit{d-d} transitions within the Cr$^{3+}$ ions.

CrPS$_4$ belongs to a broad class of transition metal phosphorus chalcogenides (TMPC), many of which, including CrPS$_4$, have been identified as S-vdW-AFs and have been extensively studied in recent years~\cite{Lee2017, Budniak2020, Alcantara2023, Gu2019, Kim2022p, Riesner2022, multian2025brightened, jana2025strong, jana2025deconstruction, Fas2025Direct, hu2025magnetic}. The selection of both the transition metal ion and the ligands determines the crystallographic structure, the exchange interactions between magnetic ions, and the anisotropy parameters, leading to diverse magnetic orderings in these compounds~\cite{lanccon2016, Calder2020, peng2020magnetic, calder2021, Wildes2017, Wildes2022}. Figure~\ref{fig:Fig1}a illustrates the widely accepted crystal structure and spin ordering of Cr$^{3+}$ ions in CrPS$_4$~\cite{Momma2011}, which emerges below the N\'eel temperature, T$_N = 38$~K~\cite{Pei2016, Calder2020}. It has a monoclinic structure with the \textit{C2} space group~\cite{Diehl1977}. Cr$^{3+}$ magnetic moments are aligned ferromagnetically within each layer along the crystal’s \textit{c}-axis, albeit with a slight tilt of approximately $12.5^{\circ}$ along the \textit{a}-axis~\cite{Calder2020}. The adjacent layers are antiferromagnetically coupled, resulting in an \textit{A}-type antiferromagnetic ordering~\cite{zhuang2016density}. 

\section{Results and discussion}

\subsection{Optical response of $\mathrm{CrPS_4}$ near low energy absorption threshold}

\begin{figure*}[b]
	\includegraphics[width=16.8cm]{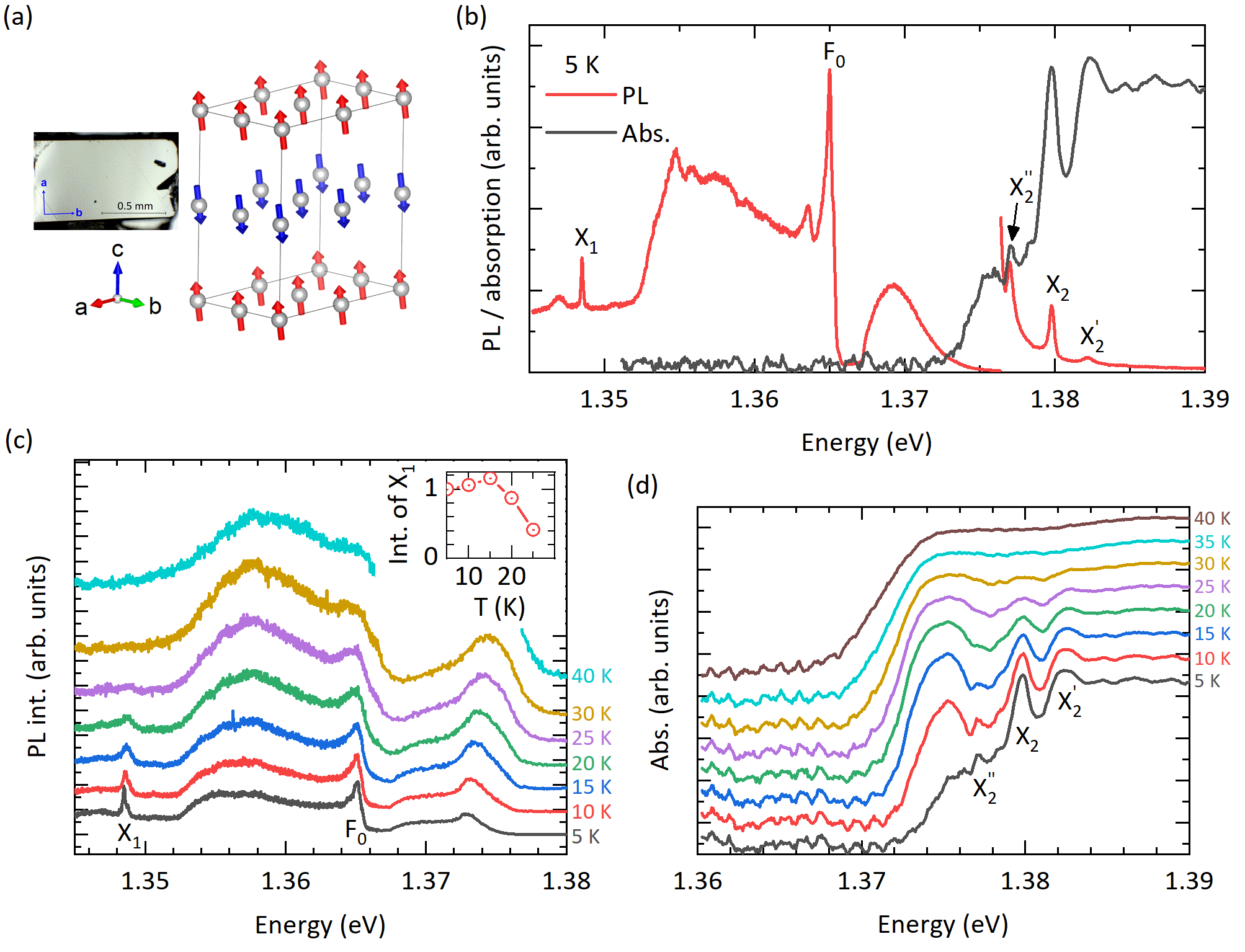}
	\caption{(a) The magnetic structure of CrPS$_4$ in antiferromagnetic phase. Grey spheres represent Cr$^{3+}$ ions while red and blue arrows represent spin orientations in adjacent layers. The figure is created using the VESTA software package. \cite{ Momma2011}. An optical image of the CrPS$_4$ crystal used in the experiments is shown on the left. (b) Near-infrared photoluminescence (red traces) and absorption (black trace) spectra of CrPS$_4$ crystal measured at 5 K without resolving the polarization of the emitted/absorbed light. The low energy PL spectrum ($\mathcal{E}~<$~1.376~eV) was measured under weak excitation intensity ($\approx$~0.25~$\mu W/ \mu m^2$), whereas significantly higher power ($\approx$~100~$\mu W/ \mu m^2$) was used for high energy PL spectrum. The transitions marked as X$_1$ and X$_2$ are of particular interest in our study. A series of (c) photoluminescence and (d) absorption spectra of CrPS$_4$ crystal, measured in E $\parallel$ \textit{b}-axis polarization, at different temperatures in the range $5$~K–$40$~K. The integrated PL intensity of X$_1$ transition is shown in the inset of Fig.~\ref{fig:Fig1}c.}
	\label{fig:Fig1}
\end{figure*}

The characteristic optical response of our CrPS$_4$ crystals, measured in the vicinity of the low-energy absorption onset and in the absence of an external magnetic field, is presented in Fig.~\ref{fig:Fig1}b-\ref{fig:Fig1}d. At low temperatures, the PL spectrum consists of a broad emission band on top of which several relatively narrow spectral features are superimposed. Among them, a prominent feature labeled F$_0$ is clearly visible. This feature has previously been discussed in terms of a Fano-type resonance~\cite{Riesner2022}. In the present work, we concentrate on the properties of the additional narrow spectral features, in particular on two well-resolved transitions labeled X$_1$ and X$_2$. The lower-energy X$_1$ resonance is observed exclusively in the PL spectra, whereas the X$_2$ transition, appearing in the high-energy tail of the PL emission, has a clear counterpart in the absorption spectrum. Detecting the X$_1$ and X$_2$ transitions is nontrivial and requires carefully controlled experimental conditions. As discussed in Section S1 of the Supporting Information (SI)~\cite{SuppInfo}, the optical response of CrPS$_4$ follows certain linear polarization selection rules. At zero magnetic field, both transitions are most readily observed when selecting linear polarization of the emitted light aligned along the crystal \textit{b}-axis ($E \parallel b$). Further, the low-power excitation is crucial for detecting the X$_1$ emission, while the X$_2$ transition appears in emission at higher excitation powers (see Fig~S1b of SI~\cite{SuppInfo}). Alternatively, the X$_2$ transition can be more suitably traced via absorption measurements; however, its very low oscillator strength necessitates the use of a thick sample. Such polarization resolved measurements were thus employed to trace the evolution of these transitions as a function of temperature, see Fig.~\ref{fig:Fig1}c~and~\ref{fig:Fig1}d. The presented PL spectra encompass the X$_1$ transition, while the absorption traces highlight the behavior of the X$_2$ resonance. Importantly, we observe that the X$_1$ and X$_2$ transitions persist below the N\'eel temperature. This is a primary indication that the X$_1$ and X$_2$ resonances are coupled to spins, as they emerge only in the spin-ordered antiferromagnetic phase. 

We also observe the emergence of several other peak-like features at low temperatures, which gradually diminish into smoother spectra as the temperature increases. In particular, the X$_1$ emission does not necessarily correspond to a single PL line, and two well-resolved transitions (X$_2'$ and X$_2''$) are discernible on the high- and low-energy sides of the X$_2$ resonance, as can be identified in Figs.~\ref{fig:Fig1}c~and~\ref{fig:Fig1}d. Moreover, the absorption onset at low temperatures is not a simple monotonic edge. It exhibits a relatively broad, peak-like structure located below the X$_2$ resonance, which is especially pronounced at intermediate temperatures. In the following, we examine the behavior of the X$_1$ and X$_2$ transitions under an applied magnetic field. The discussion about the additional spectral features introduced above is deferred to a later section.

\subsection{Spin-entangled transitions in an external magnetic field}

Compelling evidence for the spin entanglement of the X$_1$ and X$_2$ transitions arises from experiments where the application of an external magnetic field, $B$, is used to alter the alignment of the spins. The majority of the collected data concerns the behavior of the X$_1$ transition, which is, however, largely mirrored in the behavior of the X$_2$ resonance. By applying the magnetic field in different orientations, we gather information about the direction of spin alignment and trace the distinct spin phases expected to be induced by the external magnetic field. 

\begin{figure*}[htp]
\centering
\includegraphics[width=16.8cm]{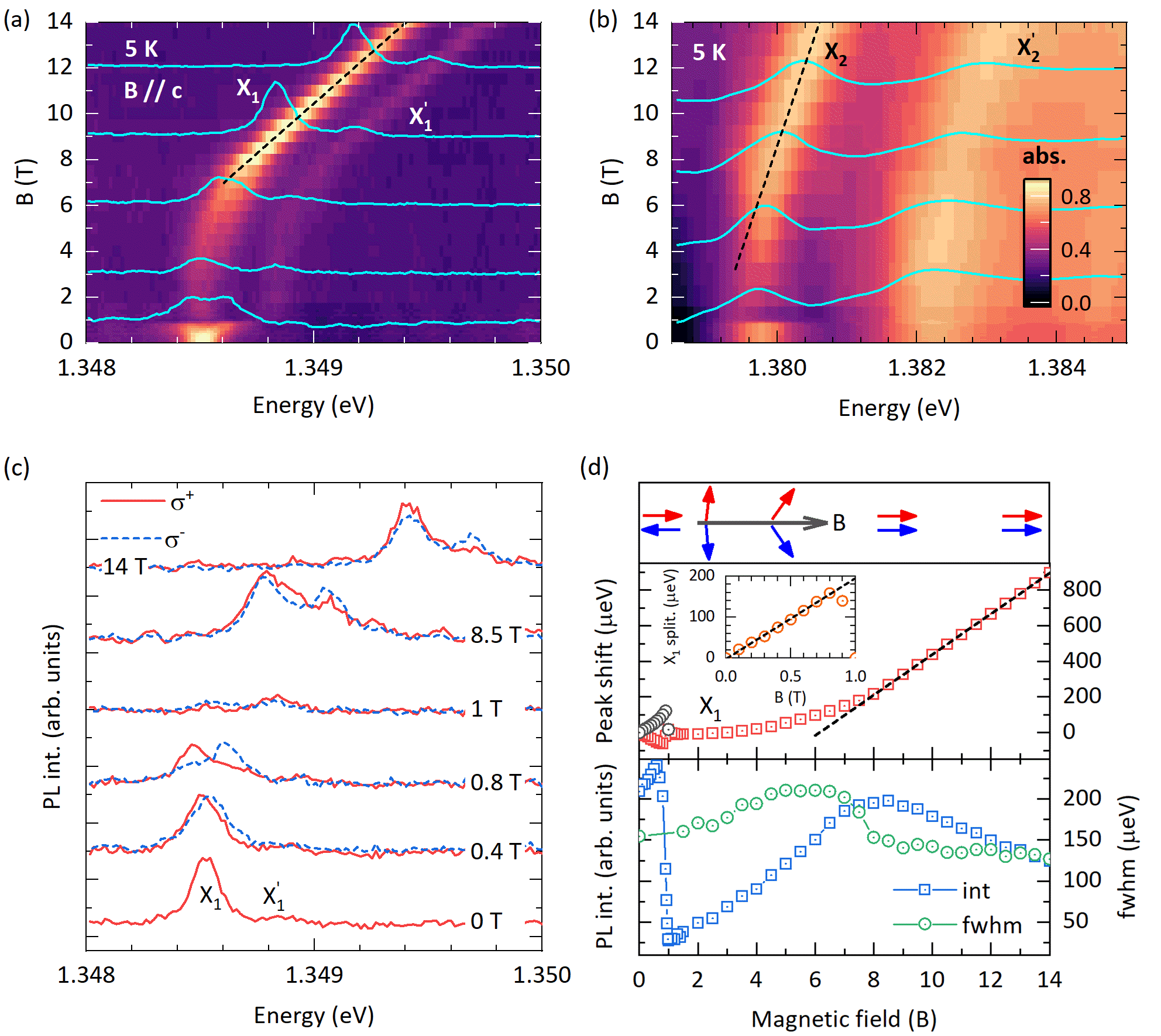}
\caption{False color map of low temperature (5~K) (a) photoluminescence (b) and absorption of CrPS$_4$ crystal as a function of the magnetic field applied along the crystal \textit{c}-axis (out-of-plane). A few representative spectra are also shown. The linear shift of the X$_1$ and X$_2$ transition above $8$~T is shown by the dashed line. (c) Circular polarization resolved PL spectra highlighting the polarization-resolved splitting of X$_1$ transition at some selected magnetic field applied along the crystal \textit{c}-axis. (d) Top: Schematic of the magnetic field-dependent spin orientation. Middle: Peak shift of X$_1$ transition with respect to $B=0~T$ peak energy as a function of the magnetic field. The dashed line corresponds to a linear fit of the magnetic field dependence above $B=8$~T. The inset shows the X$_1$ energy splitting for B$<$1~T. Bottom: Integrated PL intensity and full width at half maxima (fwhm) of the X$_1$ transition as a function of magnetic field.}
\label{fig:Fig2}
\end{figure*}

High-resolution PL spectra focused on the region of the X$_1$ transition for a magnetic field ($B$) applied perpendicular to the layers (along the \textit{c}-crystal axis) are presented in Fig.~\ref{fig:Fig2}a. The field evolution of the X$_2$ resonance is traced with absorption measurements and shown in Fig.~\ref{fig:Fig2}b. The PL measurements were performed at low excitation power without resolving the polarization of the emitted light. The X$_1$ transition splits progressively into two components as the magnetic field increases up to $\approx0.9$~T. This splitting is more clearly observed through circular polarization-resolved PL measurements (see Fig.~\ref{fig:Fig2}c), where the two components exhibit partial circular polarization with opposite helicity. A notable suppression of the X$_1$ emission occurs around $1$~T. As the magnetic field increases beyond $1$~T, the two X$_1$ components merge into a single PL line which gains in intensity and shifts linearly towards higher energy (blue shift)  at  $B \geq 8$~T. The weak X$_{1}^{'}$ satellite largely follows the evolution of the X$_1$ transition, though it is barely pronounced at low magnetic fields, as shown in Fig.~\ref{fig:Fig2}a and \ref{fig:Fig2}c. The splitting of the X$_2$ resonance below $1$~T is not observed in the absorption spectra due to its relatively broad linewidth. However, it shows a pronounced intensity suppression at $1$~T and a linear magnetic field dependence for $B \geq 8$~T, similar to those of the X$_1$ transition. The less pronounced X$_2^{'}$ satellite, appearing on the higher energy side of the X$_2$ transition, also shows an identical magnetic field-dependent blueshift. Such evolution of the X$_1$ and X$_2$ transitions with the magnetic field strongly supports their spin-entangled character. Similar spin-entangled excitons have been reported in other S-vdW-AFs (MnPS$_3$ and NiPS$_3$)~\cite{Gnatchenko2011, jana2025deconstruction, Dipankar2023}.

The X$_1$ transition is sensitive to the apparent configuration of the spin ordering (see upper panel of Fig.~\ref{fig:Fig2}d). To quantify the magnetic field evolution of the X$_1$ emission (as shown in Fig.~\ref{fig:Fig2}a), we modeled the X$_1$ transition using two Lorentzian functions with equal half-width and intensity. These Lorentzian components overlap perfectly at $B=0$~T and again when $B > 1$~T. With the results of the simulations, we retrieve the initial splitting of the X$_1$ transition, its collapse into a single line around $1$~T, and its progressive blue-shift at higher fields (see middle panel of Fig.~\ref{fig:Fig2}d). Following earlier studies on spin-entangled transitions~\cite{jana2025deconstruction, Dipankar2023, Wang2024}, we assume that the magnetic-field evolution of the X$_1$ transition for a field applied along the \textit{c}-axis -- roughly corresponding to the direction of the zero-field spin alignment -- can be described by the following spin-sensitive formula:

\begin{equation} \label{eq:1}
\mathcal{E}_{X_1}^{\uparrow,\downarrow}(B)=\mathcal{E}_{X_1}-g\mu_B B \cos\Psi ^{\uparrow,\downarrow}(B)
\end{equation}

where $\Psi^{\uparrow,\downarrow}$ are the apparent angles between the spin’s direction and the applied magnetic field for the two spin sublattices, $\mu_B$ is the Bohr magneton, and $g$ stands for the effective $g$-factor.  At low magnetic fields, within the antiferromagnetic phase, all spins remain aligned along the $B$-direction, with spins from one sublattice oriented parallel to $B$ ($\Psi ^{\uparrow}=0^{\circ}$), while those from the second sublattice are oriented antiparallel to $B$ ($\Psi ^{\downarrow}=180^{\circ}$). The Eq.~\ref{eq:1} implies a splitting of the X$_1$ resonance into two components. The splitting value between the two components, $\Delta \mathcal{E} (B)= 2|g|\mu_B B$ is linear with the magnetic field, which is indeed seen in the experimental data (inset in the middle panel of Fig.~\ref{fig:Fig2}d), from which we estimate $|g|\approx1.7$. According to Eq.~\ref{eq:1} the two-component X$_1$ transition merges into a single line when $\Psi ^{\uparrow} = \Psi ^{\downarrow} = 90^{\circ}$. Such conditions are met at higher magnetic fields (see upper panel of Fig.~\ref{fig:Fig2}d), above the characteristic threshold of the spin-flop field ($B_{flop}$). Above $B_{flop}$, the magnetic moments switch to another equilibrium configuration in which the magnetic moments are aligned perpendicular to the magnetic field. This is reflected in the experimental data, and in accordance with the previous report~\cite{wu2023gate, li2023ultrastrong, Pei2016, peng2020magnetic}, the value of $B_{flop}=0.9 \pm 0.1$~T is derived for the CrPS$_4$ antiferromagnet. Upon further increase of the magnetic field, the system enters a canted phase in which spins in both sublattices progressively rotate toward the field direction, eventually reaching a fully ferromagnetic alignment above a characteristic spin-saturation field, which we denote as $B_{sat}$. For $B>B_{sat}$, both angles $\Psi ^{\uparrow}$ and $\Psi ^{\downarrow}$ approach $0^{\circ}$  and Eq.~\ref{eq:1} yields a linear magnetic-field dependence of the X$_1$ transition energy, proportional to $-g\mu_B B$. Such a linear behavior is indeed observed experimentally for fields exceeding the estimated $B_s=8$~T. In this regime, we extract an effective \textit{g}-factor of $g=-1.97$, where, consistent with Eq.~\ref{eq:1}, the negative sign reflects the experimentally observed increase of the transition energy with magnetic field. We note that the \textit{g}-factor estimated from the high-field data differs slightly from that extracted from the low-field splitting of the X$_1$ transition. This discrepancy is most likely due to the limited precision in determining the separation between the X$_1$ components at low magnetic fields. The prominent characteristics of the X$_1$ transition, encompassing the splitting into two components in the low-field antiferromagnetic phase, the collapse into a single line above $B_{flop}$, and the linear field dependence at high magnetic fields above $B_{sat}$, are thus well described by Eq.~\ref{eq:1}. Nevertheless, the extracted negative sign of the effective g factor, which is opposite to that reported for spin-entangled transitions in the antiferromagnet MnPS$_3$~\cite{jana2025deconstruction}, constitutes a differentiating signature of the CrPS$_4$ crystals.

Less clear is the energy evolution of the X$_1$ transition in the intermediate magnetic-field range $B_{flop} < B < B_{sat}$, corresponding to the spin-canted phase. Within a simple model, the spin orientations in the two sublattices satisfy $\Psi ^{\uparrow}= -\Psi ^{\downarrow}$ and progressively rotate toward the field direction, approaching $0^{\circ}$ at $B_{sat}$. The exchange energies in CrPS$_4$ largely exceed the anisotropy energies~\cite{Calder2020}, therefore we expect that in this regime: $\cos(\Psi ^{\uparrow,\downarrow})=g\mu_B B / g \mu_B B_{sat}$~\cite{cho2023}. Under such conditions, Eq.~\ref{eq:1} implies a quadratic dependence of the X$_1$ transition energy on the magnetic field~\cite{jana2025deconstruction}. As illustrated in Fig.~\ref{fig:Fig2}d, we observe a blue shift of the X$_1$ transition with the magnetic field (along the \textit{c}-axis) till $B_{sat}$, but the shift is considerably smaller than the estimated value ($\Delta \mathcal{E} (B_{sat}) = g \mu_B B_{sat}=0.92$~meV) based on Eq.~\ref{eq:1}. The spin-canted phase introduces additional complexity due to the noncollinear arrangement of spins, complicating the interpretation of the magnetic field's influence on the X$_1$ transition. This phase likely involves a finite degree of spin disorder in the spin-canted phase, as evidenced by the noticeable broadening of the X$_1$ transition when $B_{flop} < B < B_{sat}$ (see Fig.~\ref{fig:Fig2}d, lower panel)~\cite{Wang2024}.

The intensity of the X$_1$ (and X$_2$) transition constitutes an alternative parameter sensitive to the emergence of different magnetic phases in the CrPS$_4$ antiferromagnet. As shown in Fig.~\ref{fig:Fig2}d, an abrupt drop in X$_1$ intensity at $B_{flop}$ is followed by a linear increase in intensity within the spin-canted phase, reaching a maximum at $B_{sat}$, and then gradually decreasing as the magnetic field increases beyond the saturation threshold. The drastic suppression of the intensity of X$_1$ and X$_2$ transitions in the spin-flop phase is a remarkable and unexpected effect. This observation indicates that the spin alignment in the spin-flop phase significantly modifies the selection rules, which govern the oscillator strength of X$_1$ and X$_2$ transitions. To gain further insight into this feature, magneto-PL spectra focused on the X$_1$ resonance were measured while applying an in-plane magnetic field oriented along the crystal \textit{a}- or \textit{b}-axes ($B \parallel a$ or $B \parallel b$ configurations) and collecting the light emission along the \textit{c}-axis. The results of these experiments are shown in Fig.~\ref{fig:Fig3}a~and~\ref{fig:Fig3}b. In both configurations, the energy of the X$_1$ transition is practically field-independent in the range $B \leq B_{sat}$, similar to the canted phase in $B \parallel c$~axis configuration. With the field exceeding $B_{sat}$, the spins are aligned along the \textit{a}- or \textit{b}-crystal axes for $B \parallel a$ or $B \parallel b$ configurations, respectively. As shown in Fig.~\ref{fig:Fig3}a~and~~\ref{fig:Fig3}b, the X$_1$ transition blueshifts linearly with $B$ above $B_{sat}$ with a slope ($|g|=1.97$) identical to the case of $B \parallel c$ configuration. Focusing now on the changes in the X$_1$ intensity, we observed very distinct behaviors in the two configurations. When $B \parallel a$-axis, the X$_1$ intensity smoothly increases with the applied field. On the contrary, when $B \parallel b$-axis, the X$_1$ intensity weakens under the influence of the magnetic field, with enhanced efficiency of the decrease observable at fields above $B_{sat}$, i.e., when spins are aligned parallel to the crystallographic \textit{b}-axis. Notably, a similar reduction in X$_1$ intensity is observed above the spin-flop field in the $B \parallel c$ configuration (see Fig.~\ref{fig:Fig2}a). As the crystallographic \textit{b}-axis corresponds to the intermediate anisotropy axis~\cite{li2023ultrastrong, Calder2020}, the spins align along the \textit{b}-axis once the spin-flop threshold is exceeded. The observed decrease (increase) in X$_1$ intensity when the spins are oriented along the \textit{b}-axis (\textit{a}-axis) thus indicates that the oscillator strength of the transition is strongly dependent on the spin orientation with respect to the crystal axes. Such an on-site spin-flip transition is optically forbidden under spin and parity selection rules. In the magnetically ordered phase, however, these selection rules are partially relaxed, enabling the transition to acquire finite intensity. Our results indicate that the degree of selection-rule relaxation exhibits highly anisotropic characteristics determined by the spin orientation with respect to the crystallographic axes. This constitutes a unique type of selection rule in CrPS$_4$ crystals inherently coupled to the magnetic order, giving rise to the observed axis-dependent modulation of the spin-entangled transition intensity.

\subsection{Origin of NIR optical response and spin entangled transitions}

With the fundamental band gap of CrPS$_4$ located at relatively high energy, around $2.4$~eV~\cite{Louisy1978}, its optical response at lower energies is commonly attributed to localized optical transitions within the \textit{3d}$^3$ electronic configuration of the Cr$^{3+}$ ions. Optical transitions associated with ions in the \textit{3d}$^3$ configuration are well known and have been observed in a wide variety of solid-state systems~\cite{van1967optical, hori1979zeeman, back2020effective, sugano1958absorption}. Their interpretation is typically based on crystal-field theory, employing configurational diagrams following the Tanabe–Sugano energy-level scheme~\cite{tanabe1954absorption}. Strictly speaking, this approach has proven successful in describing optical transitions in numerous crystals in which the transition metal ions are introduced as dopants~\cite{back2020effective}, with the archetypal example being ruby ($\alpha$-Al$_2$O$_3$:Cr$^{3+}$)~\cite{sugano1958absorption, hori1979zeeman}. In such systems, the optical spectra are interpreted in terms of transitions from the ground-state quartet $^4A_2$ to excited states, including the $^2E$ and $^2T_1$ doublets, as well as the $^4T_2$ and $^4T_1$ quartets. Transitions conserving spin multiplicity (quartet-to-quartet) typically give rise to relatively strong and broad absorption and photoluminescence features, enabled by vibronic coupling. In contrast, quartet-to-doublet transitions are much weaker and appear as sharp spectral lines. Their optical response is stimulated by spin–orbit coupling and/or by lowering the local symmetry of the Cr$^{3+}$ sites.

\begin{figure}[htp]
\centering
\includegraphics[width=16.8cm]{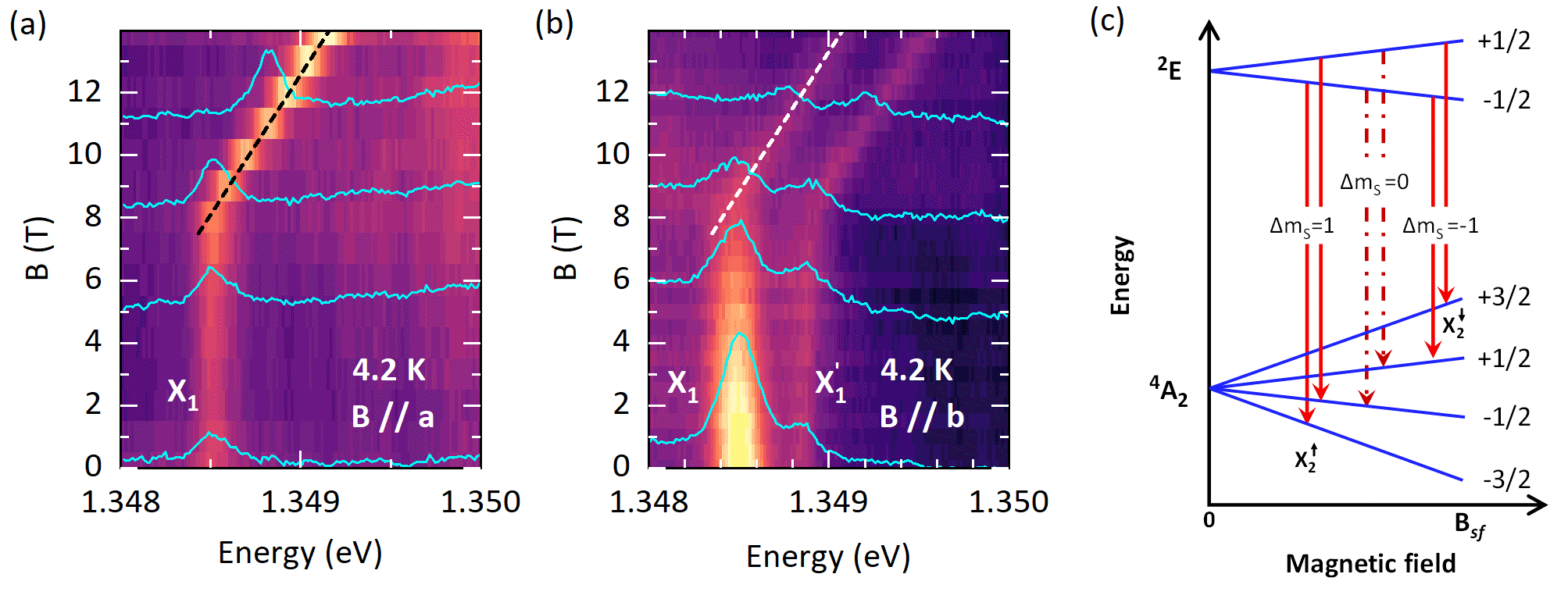}
\caption{False color map of the PL spectra of CrPS$_4$ measured at low temperature (5~K) as a function of the magnetic field applied along (a) crystal \textit{a}-axis and (b) crystal \textit{b}-axis. The light propagation vector was perpendicular to the sample plane (k $\parallel$ \textit{c}-axis) for both configurations. The dashed line above $8$~T corresponds to a linear fit to the magnetic field dependence of X$_1$ transition. (c) Schematic illustration of the transitions occurring between the $^4A_2$ ($S=3/2$) and $^2E$ ($S=1/2$) multiplets of  Cr$^{3+}$ ions in the presence of an external magnetic field.}
\label{fig:Fig3}
\end{figure}

The molecular-field model well functional for isolated Cr$^{3+}$ centers may not strictly apply when Cr$^{3+}$ ions form an integral part of the crystal lattice, as in CrPS$_4$. Worth noting in this context is a report~\cite{van1967optical} on the interpretation of the optical response of Cr$_2$O$_3$, another Cr-based antiferromagnet, in which the formation of Frenkel-type excitons originating from single-ion $^4A_2$$\rightarrow$$^2E$ transitions, and strongly influenced by exchange interactions, are invoked to account for the experimental observations. Moreover, the close energetic proximity and ordering of the excited $^2E$, $^2T_1$, and $^4T_2$ states of Cr$^{3+}$ ions significantly influence the spectral characteristics. In light of these considerations, an unambiguous interpretation of the near-infrared optical response of CrPS$_4$ remains elusive~\cite{Kim2022p, multian2025brightened, Riesner2022, Gu2019}. Here, we speculate on the possible origin of the multifaceted optical transitions, accounting for our characterization of newly observed narrow resonances. First, in agreement with previous reports, we confirm that the broad photoluminescence and absorption bands extending toward lower and higher energies, respectively, from approximately $1.375$~eV at low temperatures closely resemble the spectra associated with the $^4A_2$$\rightarrow$ $^4T_2$ transitions of isolated Cr$^{3+}$ ions. The narrower resonances superimposed on the broad spectral response can thus be naturally attributed to direct optical transitions and/or their phonon replicas associated with the $^4A_2$$\rightarrow$ $^2E$ and $^4A_2$$\rightarrow$ $^2T_1$ single-ion transitions (see section S2 of SI~\cite{SuppInfo}). The observed X$_2$ and X$_1$ resonances are of our particular interest. As illustrated in Fig.~\ref{fig:Fig3}c, we propose that X$_2$ resonance is associated with the zero-phonon transition, resembling the $^4A_2$$\rightarrow$ $^2E$ direct transitions with $\Delta m_S=\pm 1$ observed in single Cr$^{3+}$ ions. This $\Delta m_S=\pm 1$ transition exhibits a Zeeman shift that depends on the orientation of the applied magnetic field relative to the spin direction, as described by Eq.~\ref{eq:1}~\cite{Gnatchenko2011, jana2025deconstruction, van1967optical}. X$_1$, X$_2^{\prime}$, and X$_2^{\prime\prime}$ can be assigned as the phonon replica of the X$_2$ transition, manifested symmetrically on both the high- and low-energy sides of the fundamental resonance (see Fig.~S3a,b of SI~\cite{SuppInfo}). The energy separation between the X$_2$ and X$_1$ resonances is $\Delta \mathcal{E} \approx$~31.3 meV ($\approx$ 250~cm$^{-1}$) while the separation between the X$_2$ and X$_2^{\prime}$ resonances is $\Delta \mathcal{E}^{\prime} \approx 2.5$ meV ($\approx 20$ cm$^{-1}$). Phonons with matching energy are not observed to be Raman active (see Fig. S3c of SI~\cite{SuppInfo}). However, CrPS$_4$, due to a relatively complex unit cell configuration, exhibits a rich spectrum of $\Gamma$-point ($k = 0$) phonon modes, which is particularly dense in the 50–400~cm$^{-1}$ range~\cite{li2025investigation}. Within this manifold, an optical-like phonon with energies close to 250~cm$^{-1}$ at $k = 0$ can be identified, accounting for the X$_1$ phonon replica of the X$_2$ zero-phonon resonance. An alternative interpretation of X$_2^{\prime}$ and X$_2^{\prime\prime}$ is their attribution to magnon-continuum replicas of the X$_2$ resonance, as reported in other S-vdW-AFs~\cite{Gnatchenko2011, Kang2020}.
Similarly, the X$_1^{\prime}$ transition that appears just above the X$_1$ resonance can be attributed to a replica of the X$_2$ resonance arising from an adjacent phonon branch.

\section{Conclusions}

In summary, we have identified and comprehensively characterized previously unreported ultra-narrow near-infrared optical transitions in the van der Waals antiferromagnet CrPS$_4$. Through combined magneto-photoluminescence and magneto-absorption measurements, we demonstrate that these resonances are intrinsically spin-entangled excitations, whose energies and intensities are governed by the underlying magnetic order. Their magnetic-field evolution provides direct optical fingerprints of the distinct spin phases of CrPS$_4$. The linear Zeeman splitting in out-of-plane fields reflects the collinear antiferromagnetic configuration and enables the extraction of the effective g-factor. The collapse of the splitting at the spin-flop transition and the high-field linear shift above the spin-saturation threshold allow us to determine the key magnetic parameters, $B_{flop}=0.9$~T and $B_{sat}=8$~T, entirely optically. Equally importantly, the pronounced and highly anisotropic modulation of the transition intensity reveals spin-orientation-dependent selection rules, offering compelling evidence for finite in-plane anisotropy and confirming the biaxial antiferromagnetic character of CrPS$_4$. The emergence of spin-entangled optical resonances in CrPS$_4$, alongside their presence in other layered antiferromagnets (e.g., NiPS$_3$ and MnPS$_3$), calls for a unified theoretical framework to elucidate their microscopic origin. The demonstrated ability to access and distinguish magnetic phases purely optically opens a pathway toward non-contact diagnostics of spin order at the microscale, selective optical manipulation of spin configurations, and the design of novel spin-sensitive optoelectronic devices.

\section{Experimental}

\textbf{Preparation of the samples.} Bulk CrPS$_4$ crystals from two different sources were investigated: commercially available crystals obtained from HQ Graphene and crystals grown by the chemical vapor transport (CVT) method. 
Thick samples with thicknesses of approximately 200–500~$\mu$m were mounted on silicon substrates using adhesive. This thickness range was chosen to suppress interference effects arising from multiple internal reflections within the crystal while ensuring sufficient optical response from the weak sub-bandgap resonances.

\textbf{PL and absorption measurements.} A micro-optical measurement setup was employed for PL and absorption experiments. For PL measurements, the sample was excited using a continuous-wave 515.5~nm laser focused on the sample surface through a 50$\times$ microscope objective, resulting in an excitation spot size of approximately 1~$\mu$m. The emitted luminescence was collected by the same objective, spectrally dispersed using a 0.7~m monochromator, and detected with a silicon-based charge-coupled device (CCD) camera cooled to 120~K. Absorption measurements were carried out in a double-pass transmission geometry using the same micro-optical setup as for PL, with the excitation laser replaced by a broadband light source. Light from a quartz–tungsten–halogen lamp was focused onto the Si substrate (beneath the sample) through the sample using a 50$\times$ microscope objective. The light reflected from the substrate consequently traverses the sample again and is collected by the same objective, yielding the transmitted intensity (I$_{trans}$). A reference spectrum (I$_{ref}$) was obtained from the reflected signal of the bare Si substrate without the sample in the optical path. The absorption was then calculated as $I_{abs}=-log(I_{trans}/I_{ref})$. For temperature and magnetic field dependent measurements, the sample was placed either on the cold finger of the liquid helium flow cryostat (5~K to 300~K) or inside a homemade setup for magneto-optical investigations ($\pm~9$~T).

\section*{Acknowledgements}

This project was supported by the Ministry of Education (Singapore) through the Research Centre of Excellence program (grant EDUN C-33-18-279-V12, I-FIM), and Academic Research Fund Tier 2 (MOE-T2EP50122-0012). This material is based upon work supported by the Air Force Office of Scientific Research and the Office of Naval Research Global under award number FA8655-21–1-7026. M.P. acknowledges support from the CENTERA2, FENG.02.01-IP.05-T004/23 project funded within the IRA program of the FNP Poland, cofinanced by the EU FENG Programme. C.F. acknowledges support from ANR-23-QUAC-0004, from CEFIPRA project n$^\circ$ 7104-2. Z.S. was supported by project LUAUS25268 from Ministry of Education Youth and Sports (MEYS) and by the project Advanced Functional Nanorobots (reg. No. CZ.02.1.01/0.0/0.0/15-003/0000444 financed by the EFRR) and by ERC-CZ program (project LL2101) from Ministry of Education Youth and Sports (MEYS).

\bibliography{achemso-demo.bib}

\newpage

\begin{figure}
	\includegraphics[width=9.0cm]{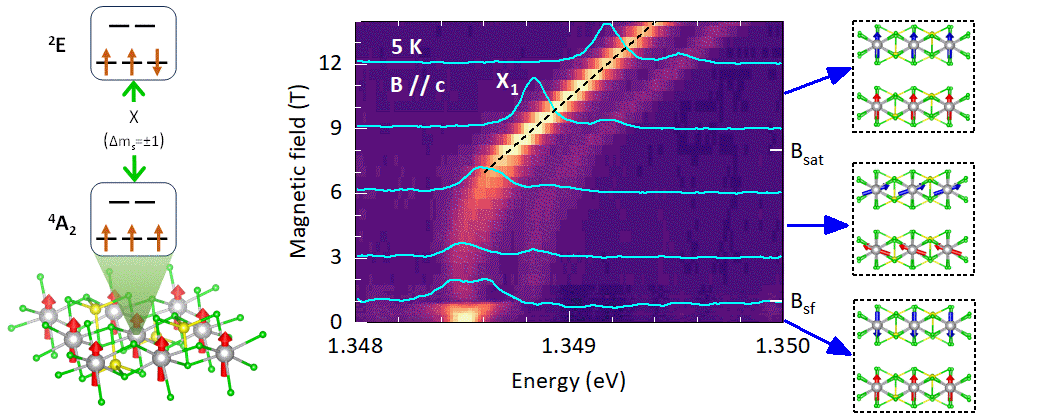}
    \caption{TOC}
\label{TOC}
\end{figure}

\end{document}